\documentclass[12pt, prd,preprintnumbers,amsmath,amssymb,superscriptaddress]{revtex4}
\usepackage{graphicx}
\usepackage{amsmath}
\usepackage{amssymb}
\usepackage{afterpage}
\usepackage{graphics}
\usepackage{float}
\usepackage{rotating}
\usepackage{xspace}
\usepackage{dcolumn}
\usepackage{bm}

\begin{document}

\title{Extracting neutrino oscillation parameters using a simultaneous fit of the $\nu_{e}$ appearance and $\nu_{\mu}$ disappearance data in the NOvA experiment\\~}

\author{{\bf Prabhjot Singh}\\
         Department of Physics and Astrophysics, University of Delhi, India \\~
        Email: prabhjot@fnal.gov \\~ \\~
        {\bf On behalf of the NOvA Collaboration} \\~ \\~
        {\bf {\small Talk presented at the APS Division of Particles and Fields Meeting (DPF 2017), July 31-August 4, 2017, Fermilab. C170731}} \\~\\~ }

\begin{abstract}
\begin{center}
{\bf Abstract\\}
\end{center}
NOvA is a two detector, long baseline neutrino oscillation experiment designed to study $\nu_{e}$ ($\bar{\nu}_{e}$) appearance
and $\nu_{\mu}$ ($\bar{\nu}_{\mu}$) disappearance in a $\nu_{\mu}$ ($\bar{\nu}_{\mu}$) beam produced at Fermilab.
The near detector (ND) is located 100 meters underground at a distance of 1 km from the target whereas the far
detector (FD) is located on the surface, 810 km away from the beam source in Ash River, MN. The ND is used to measure
the beam before oscillations and the FD measures the oscillated spectrum. The ND and the FD are functionally identical
detectors and the ND spectra are extrapolated to the FD to predict the signal and background spectra expected in the FD.
The extrapolation and data fitting techniques developed for these analyses within NOvA are presented.
\end{abstract}
\maketitle

\section{Introduction to the NOvA Experiment}
NOvA (NuMI Off-axis $\nu_{e}$ Appearance) is a two detector, long baseline, neutrino oscillation experiment.
NOvA \cite{nova_technical_report} observes neutrinos produced in Fermilab's NuMI \cite{nova_numi} (Neutrinos at Main Injector) beam line in two detectors.
The NuMI beam is composed of 97.5\% $\nu_{\mu}$, 1.8\% of ${\bar\nu}_{\mu}$ and 0.7\% of ($\nu_{e}$ + ${\bar\nu}_{e}$).
A beam of high energy, 120 GeV protons from the Main Injector impinge on a fixed graphite target producing pions and kaons \cite{nova_beam}.
Magnetic horns select pions and kaons of the desired charge and momentum and focus them into a narrow beam.
Charged pions and kaons spontaneously decay into muons and neutrinos.
240 meters of rock filters out muons produced in the decay pipe and we are left with a beam of neutrinos.
NOvA is 14.6 mrad off-axis from the NuMI beam to observe a neutrino energy spectrum peaked at 2 GeV which is
optimized for observing $\nu_{\mu} \rightarrow \nu_{e}$ oscillations.
The ND, is located 100 meters underground at a distance of 1 km from the target on-site at Fermilab and
the FD is located on the Earth's surface, 810 km away from the beam source in Ash River, MN.
NOvA is designed to study $\overset{(-)}\nu_{\mu} \rightarrow \overset{(-)}\nu_{\mu}$ disappearance,
$\overset{(-)}\nu_{\mu} \rightarrow \overset{(-)}\nu_{e}$ appearance.
Both NOvA detectors are functionally identical and constructed from planes of extruded polyvinyl chloride (PVC) cells \cite{nova_technical_report}.
NOvA cells have a rectangular cross section measuring 3.9 cm by 6.6 cm and are 15.5 m (3.9 m) long in the FD (ND).
NOvA cells are arranged in alternating horizontal and vertical planes for 3D tracking and are oriented percendicular to the NuMI beam.
There are 896 (214) planes and 344,064 (20,192) cells in the FD (ND).
Each cell is filled with liquid scintillator. Wavelenght shifting fibers are used to collect the light deposited by any charged paricle in
the detector. The fiber ends terminate on a single pixel of an avalanche photodiode (APD) \cite{nova_apd}.
The ND is used to measure the beam before oscillations and the FD measures the oscillated spectrum.
\section{Neutrino events in NOvA}
\begin{figure}[H]
  \begin{centering}
    \includegraphics[width=0.7\textwidth,height=0.30\textheight]{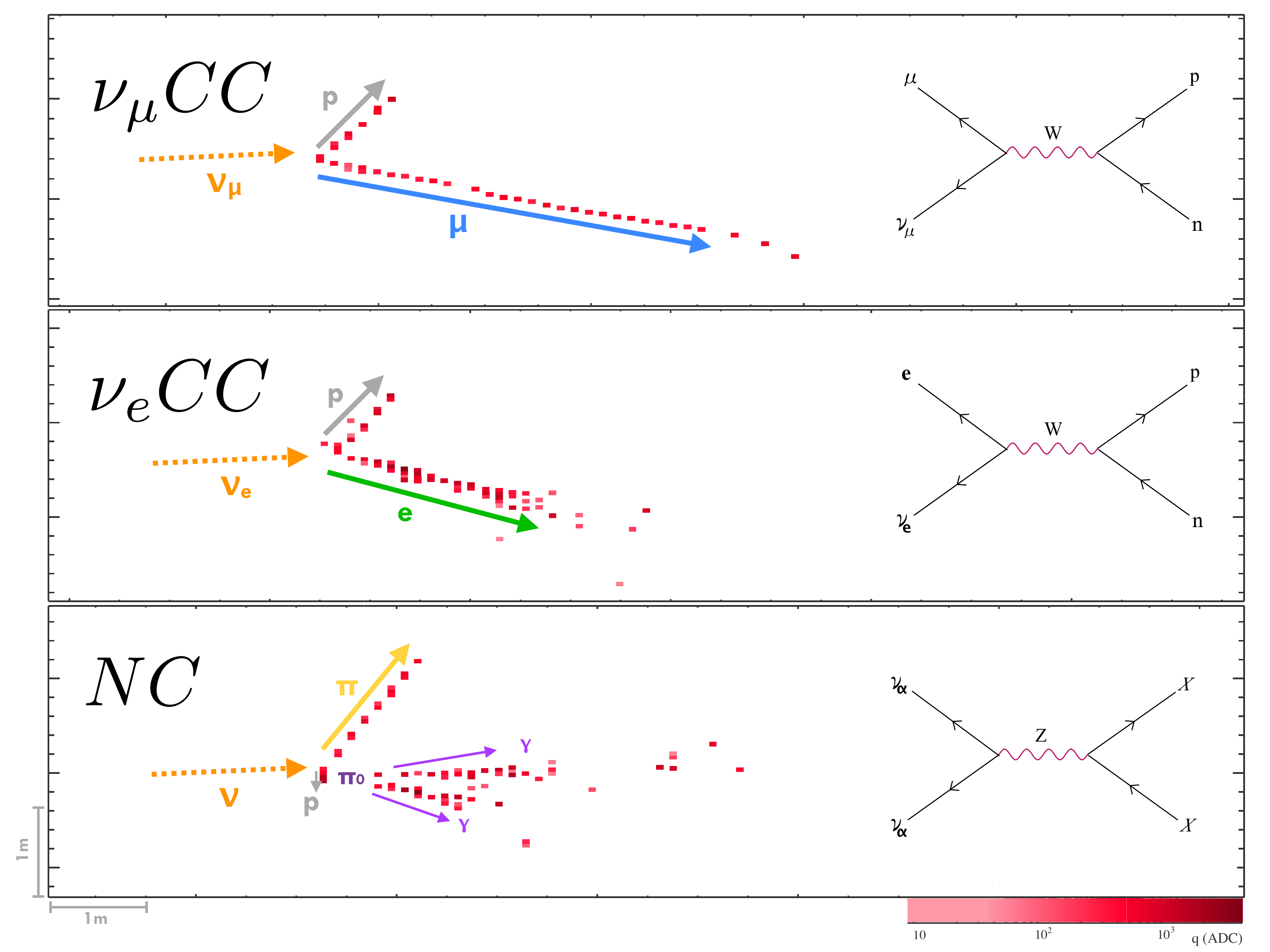}\\
  \end{centering}
  \caption{Three event topologies are observed in the NOvA oscillation analyses. $\nu_{\mu}$-CC (top) is the signal for the $\nu_{\mu}$ ($\bar{\nu}_{\mu}$) disappearance anaysis.
           $\nu_{e}$-CC (middle) is the signal for the $\nu_{e}$ ($\bar{\nu}_{e}$) appearance analysis.
           NC (bottom) is a background in $\nu_{\mu}$-disappearance and $\nu_{e}$-appearance analyses}
  \label{fig:EventTopologies}
\end{figure}

The electron neutrino charged current (CC), $\nu_{e}$-CC interaction is the signal for the $\nu_{e}$ ($\bar{\nu}_{e}$) appearance analysis.
The signature of $\nu_{e}$-CC interactions in the NOvA detectors is an electromagnetic shower plus any associated hadronic recoil energy.
The muon neutrino charged current, $\nu_{\mu}$-CC, interaction is the signal for the $\nu_{\mu}$ ($\bar{\nu}_{\mu}$) disappearance anaysis.
The signature of $\nu_{\mu}$-CC interactions in the NOvA detectors is a long track of muons and any associated hadronic activity at the vertex.
The largest background in both $\nu_{e}$-appearance and $\nu_{\mu}$-disappearance anayses arises from neutral current (NC) interactions of beam neutrinos.
The signatures of $\nu_{\mu}$ CC, $\nu_{e}$ CC and NC interactions are shown in Fig. \ref{fig:EventTopologies}.
\section{FAR DETECTOR PREDICTION USING NEAR DETECTOR}
\begin{figure}[H]
  \begin{centering}
    \includegraphics[width=0.8\textwidth,height=0.30\textheight]{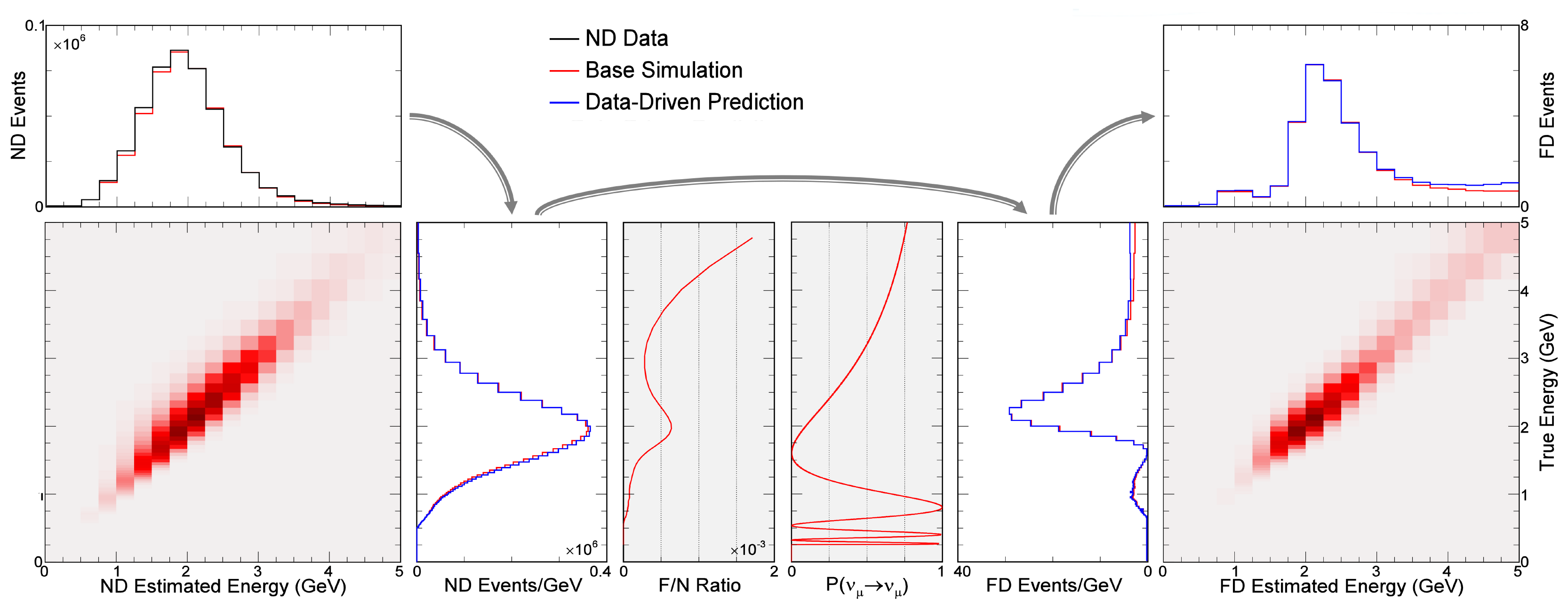}\\
  \end{centering}
  \caption{The ND spectra are used to predict the signal and backgrounds in the FD. The ND reconstructed energy spectrum is converted to the true energy spectrum.
           The FD-to-ND MC event ratio along with choice of the oscillation parameters are used to make prediction in the FD.}
  \label{fig:Extrapolation}
\end{figure}

\begin{figure}
  \begin{centering}
    \includegraphics[width=0.6\textwidth,height=0.30\textheight]{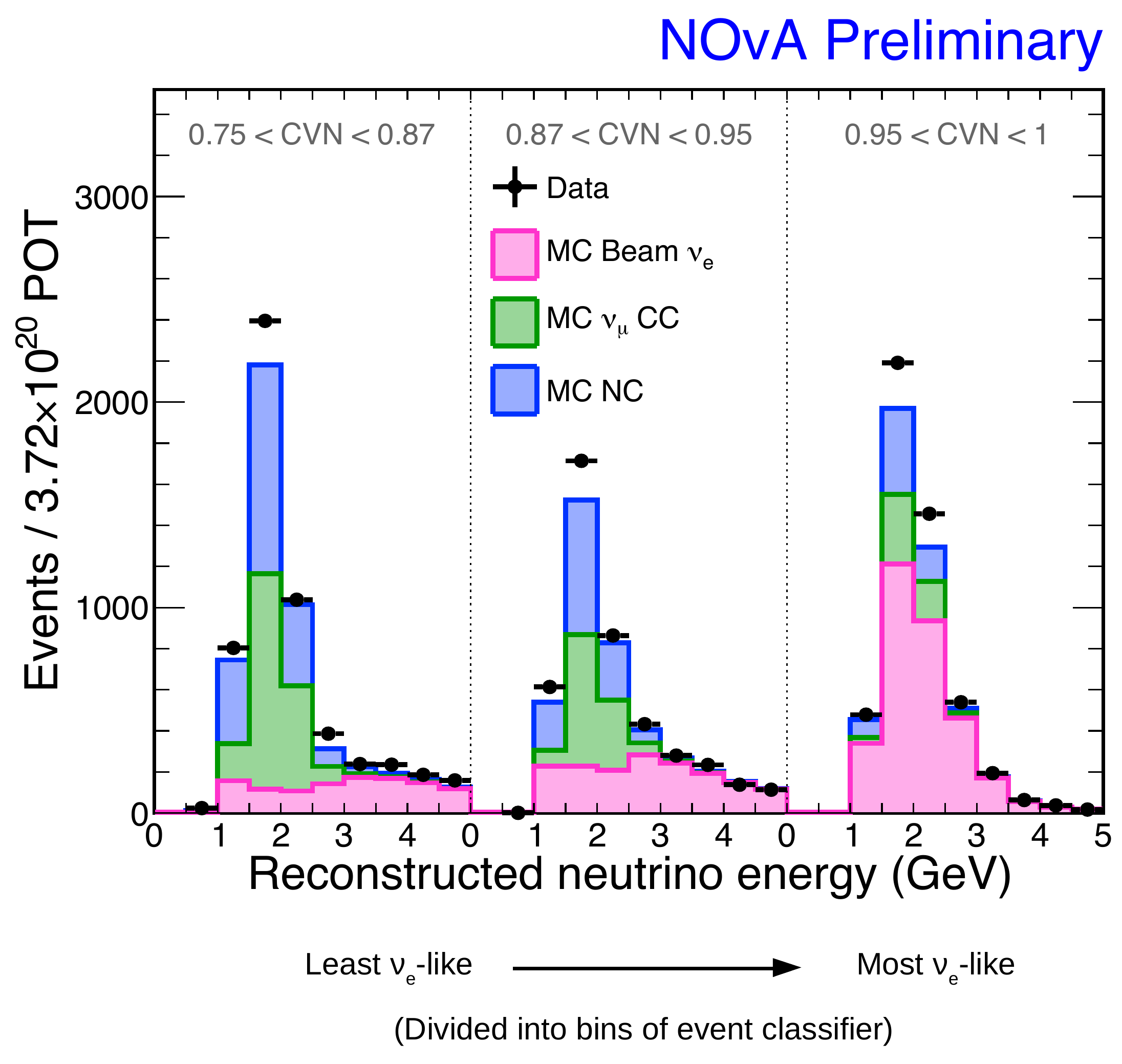}\\
  \end{centering}
  \caption{{\small In the $\nu_{e}$-appearance analysis deep learning is used to seperate the signal and backgrounds into 3 bins of identification confidence
from least $\nu_{e}$-like to the most $\nu_{e}$-like.}}
  \label{fig:Decomposition}
\end{figure}

The ND spectra are used to predict signal and backgrounds in the FD.
Both $\nu_{\mu}$ $\rightarrow$ $\nu_{\mu}$ disappearance and $\nu_{\mu}$ $\rightarrow$ $\nu_{e}$ appearance analyses use muon neutrinos in the ND for
signal prediction in the FD.
Discrepancies between data and MC calculations in the ND energy spectrum are extrapolated to produce a predicted FD spectrum.
We first convert the ND reconstructed energy spectrum into a true energy spectrum using the reconstructed-to-true migration matrix
obtained from the ND simulation, and then multiply by the FD MC to ND MC event ratio as a function of true neutrino energy to obtain
the FD true energy spectrum. The ratio also incorporates the effect of three-flavor neutrino oscillations, including matter effects,
for any particular choice of the oscillation parameters. The FD true energy prediction is transformed into a reconstructed energy
prediction using the simulated FD migration matrix. In the final step, the data-based cosmic and simulation-based beam-induced backgrounds
are added to the prediction, which is then compared to the FD data. The whole process of the extrapolation from the ND to the FD is shown in Fig. \ref{fig:Extrapolation}.
In the $\nu_{e}$-appearance analysis deep learning \cite{cvn_nue, nova_cvn} is used to seperate the signal and backgrounds into 3 bins of identification confidence
from least $\nu_{e}$-like to the most $\nu_{e}$-like as shown in Fig. \ref{fig:Decomposition}.
The $\nu_{e}$ selected beam $\nu_{e}$ CC, $\nu_{\mu}$ CC and NC interactions in the ND are the major backgrounds in the FD for the $\nu_{e}$-appearance analysis.
Since the NC, $\nu_{\mu}$ CC, and beam $\nu_{e}$ CC background components are affected differently by oscillations,
the total background selected in the ND data is broken down into these components,
then each component is extrapolated from the ND to the FD for background predictions in the FD.
\section{SYSTEMATIC UNCERTAINTIES IN THE $\nu_{\mu}$ DISAPPEARANCE AND $\nu_{e}$ APPEARANCE ANALYSES}
\begin{figure}
  \begin{centering}
    \includegraphics[width=0.8\textwidth,height=0.30\textheight]{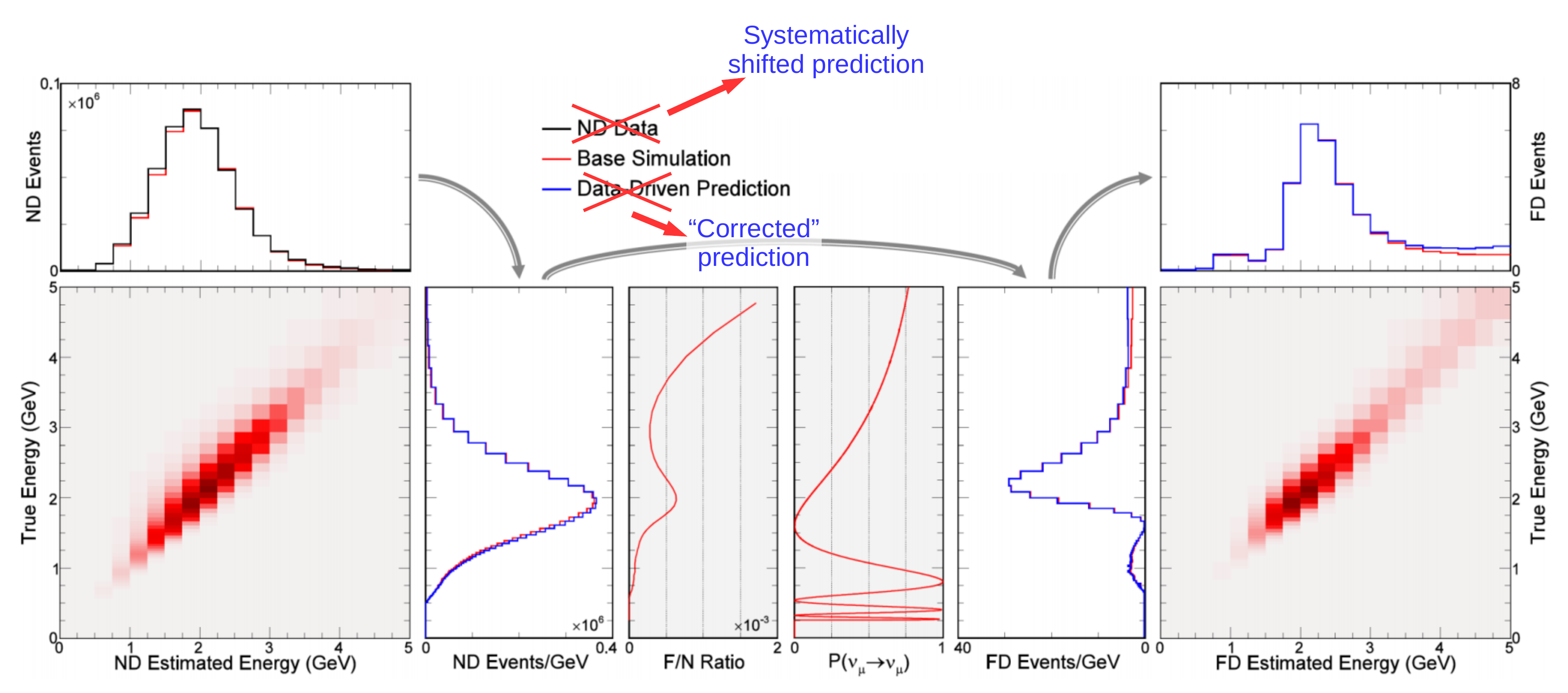}\\
  \end{centering}
  \caption{Systematic uncertainties in the analysis are extrapolated from the ND to the FD using the extrapolation technique.
           The ND Data is replaced by systematically shifted ND MC under a systematic shift.}
  \label{fig:Extrapolation_systematics}
\end{figure}

Systematic uncertainties are evaluated by reweighting or generating new simulated event samples modified to account for each uncertainty in the ND and FD.
Systematic uncertainties in the analysis are extrapolated from the ND to the FD using the same extrapolation technique.
The ND Data is replaced by systematically shifted ND MC under a systematic shift.
The corrected ND true energy spectra is then extrapolated to the FD as shown in Fig. \ref{fig:Extrapolation_systematics}.
Systematic uncertainties are included as nuisance parameters in the fit for all analyses.
For the simultaneous fit of the $\nu_{e}$-appearance and $\nu_{\mu}$-disappearance data the nuisance parameters
associated with the systematic uncertainties which are common between the two data sets, are correlated appropriately.
The functionally identical detectors allow most uncertainties to cancel when predicting the FD spectrum.
\begin{figure}[H]
\begin{minipage}[l]{0.48\textwidth}
\includegraphics[width=1.0\textwidth,height=0.20\textheight] {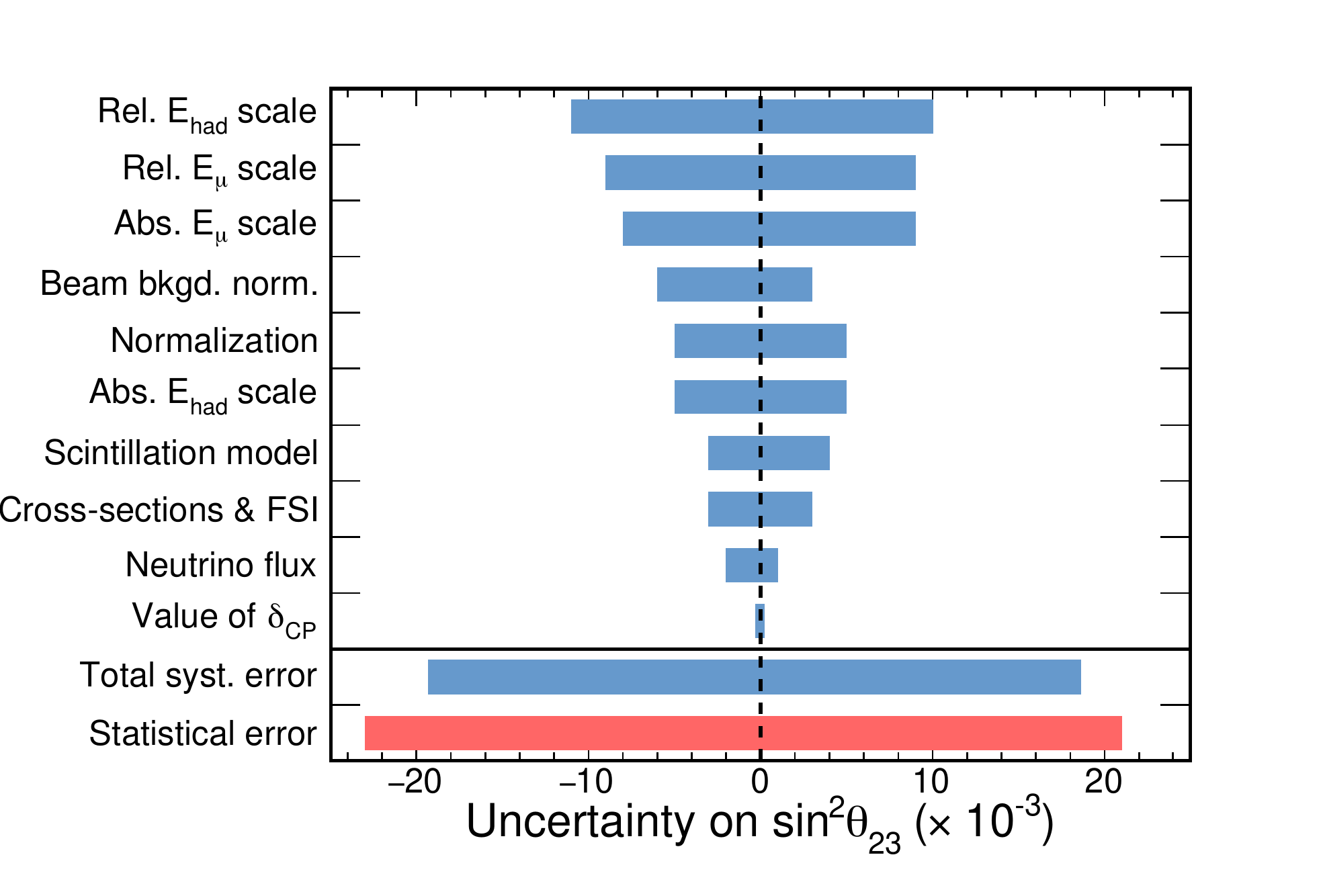}
\end{minipage}
\begin{minipage}[l]{0.48\textwidth}
\includegraphics[width=1.0\textwidth,height=0.20\textheight] {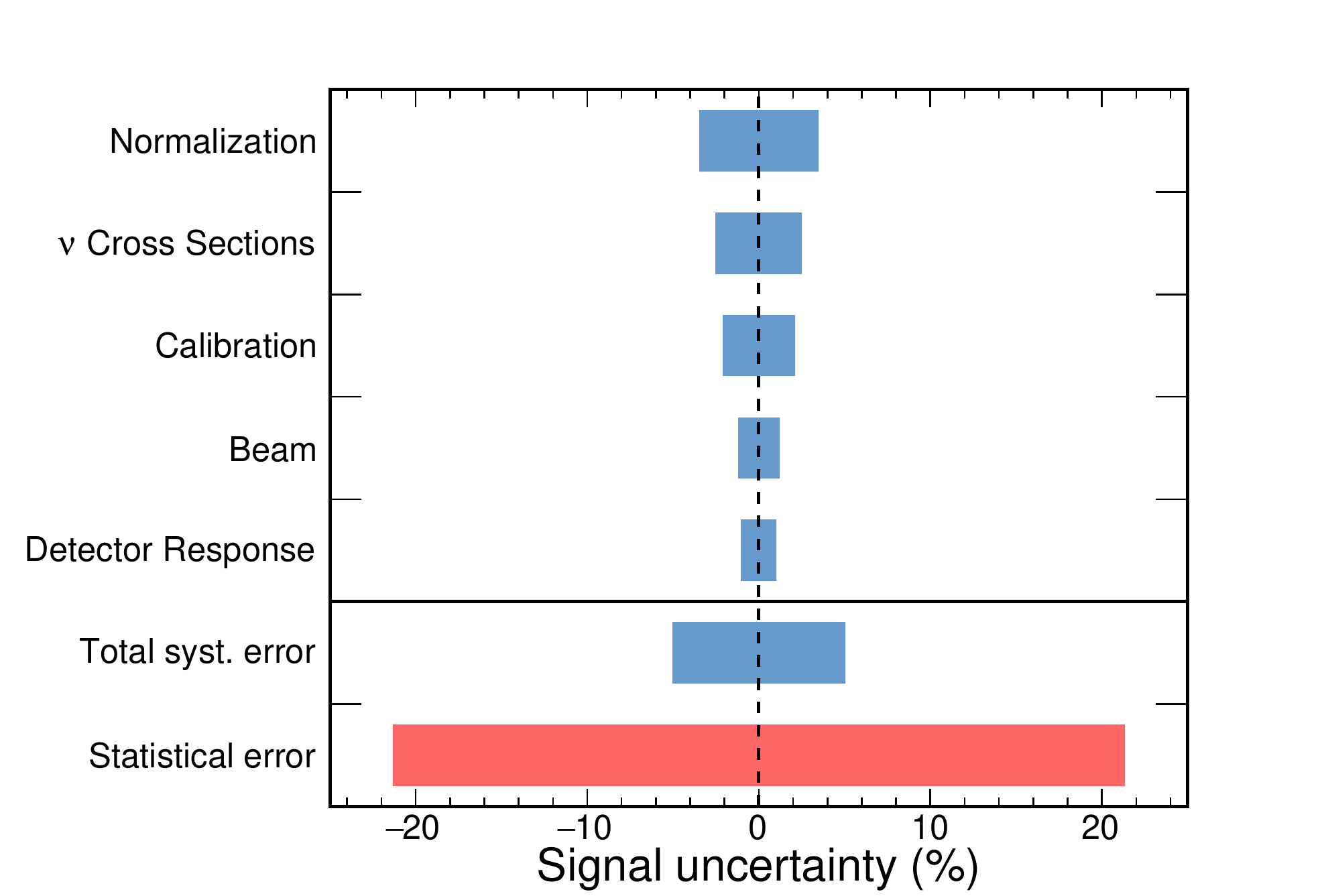}
\end{minipage}
\caption{{\small Variour sources of systematic uncertainties in the $\nu_{\mu}$ (left) and  $\nu_{e}$ (right) analyses are shown.
          In the $\nu_{\mu}$-disappearance analysis uncertainties due to detector response, i.e. relative and absolute muon and hadronic energy scales, are dominant.
          In the $\nu_{e}$-appearance analysis, cross section uncertainties are important for the signal prediction.
          Both  $\nu_{\mu}$ and $\nu_{e}$ uncertainties are small compared to statistical uncertainties.}}
  \label{fig:nue_numu_systematics}
\end{figure}
The $\nu_{\mu}$-disappearance analysis accounts for systematic uncertainties in the energy scale, normalization,
 neutrino cross section and final state interactions, neutrino flux, and backgrounds. The uncertainties due to detector response,
 that is, relative and absolute muon and hadronic energy scales are dominant but smaller than the statistical uncertainties.
In the $\nu_{e}$-appearance analysis, cross section uncertainties are important for the signal prediction, but small compared to statistical uncertainties.
\section{RESULTS}
\subsection{$\nu_{\mu}$ FIT TO THE DISAPPEARANCE DATA}
\begin{figure}[H]
\begin{center}
  \includegraphics[width=0.3\linewidth]{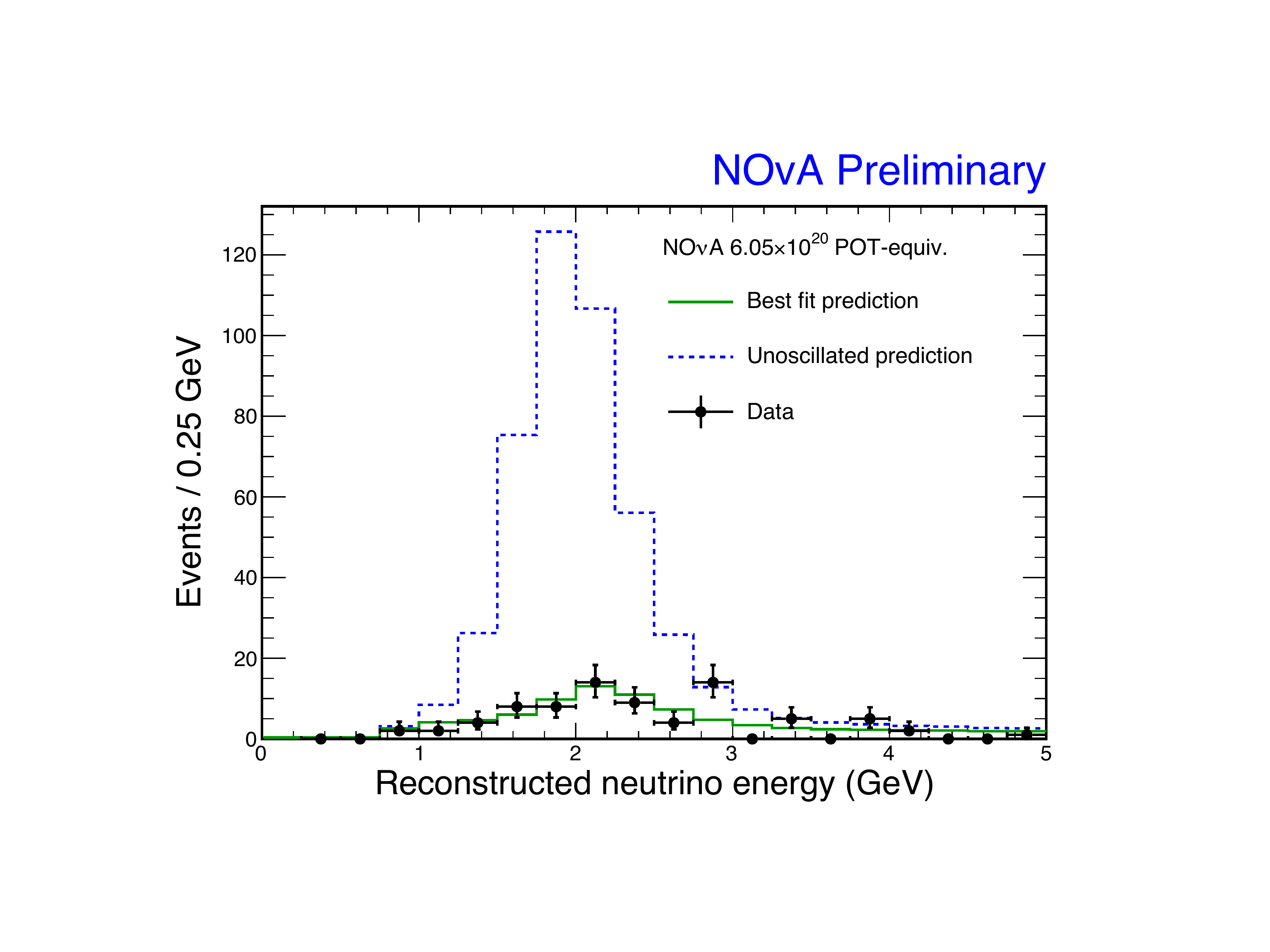}
  \includegraphics[width=0.3\linewidth]{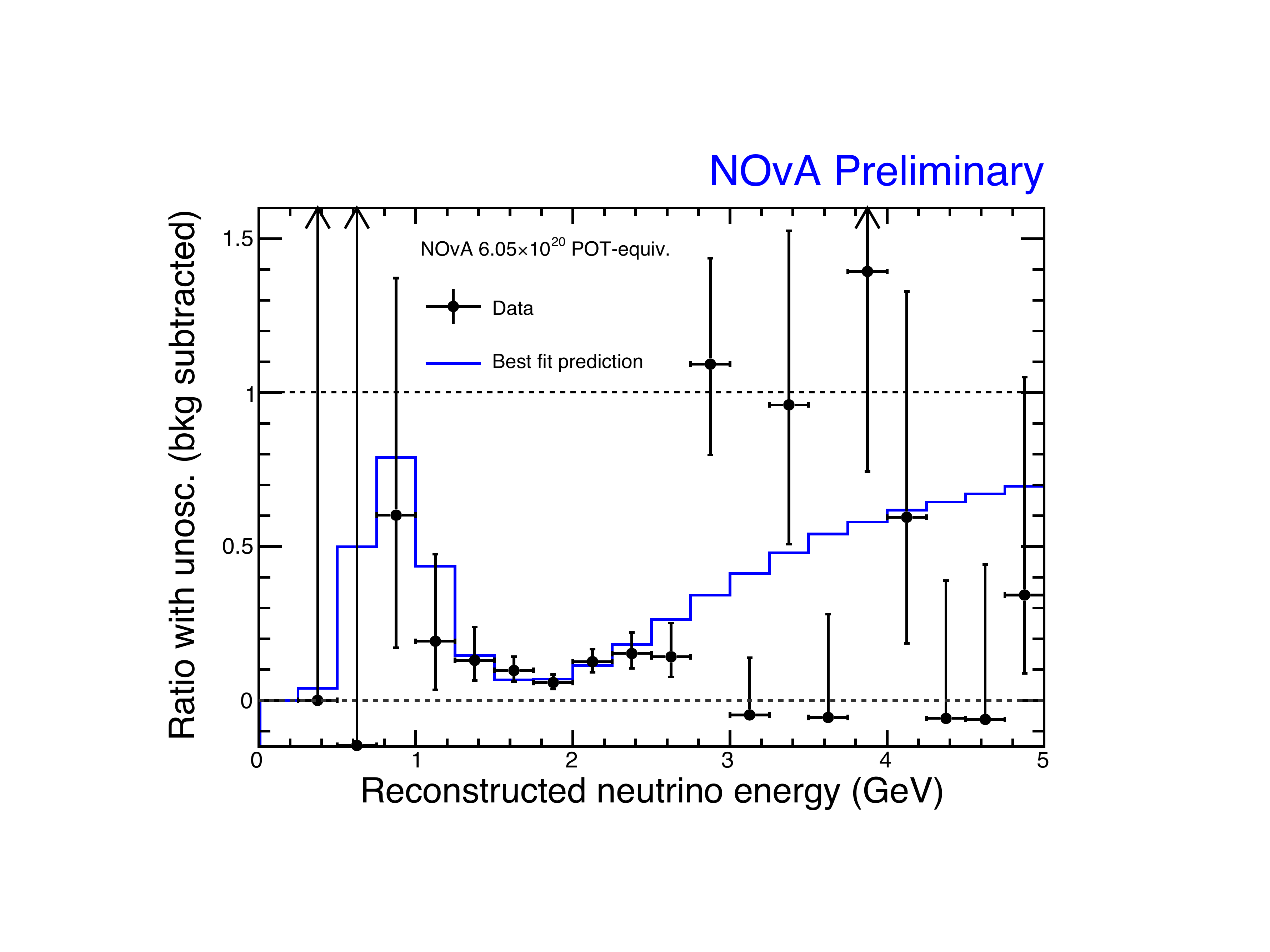}
  \includegraphics[width=0.3\linewidth]{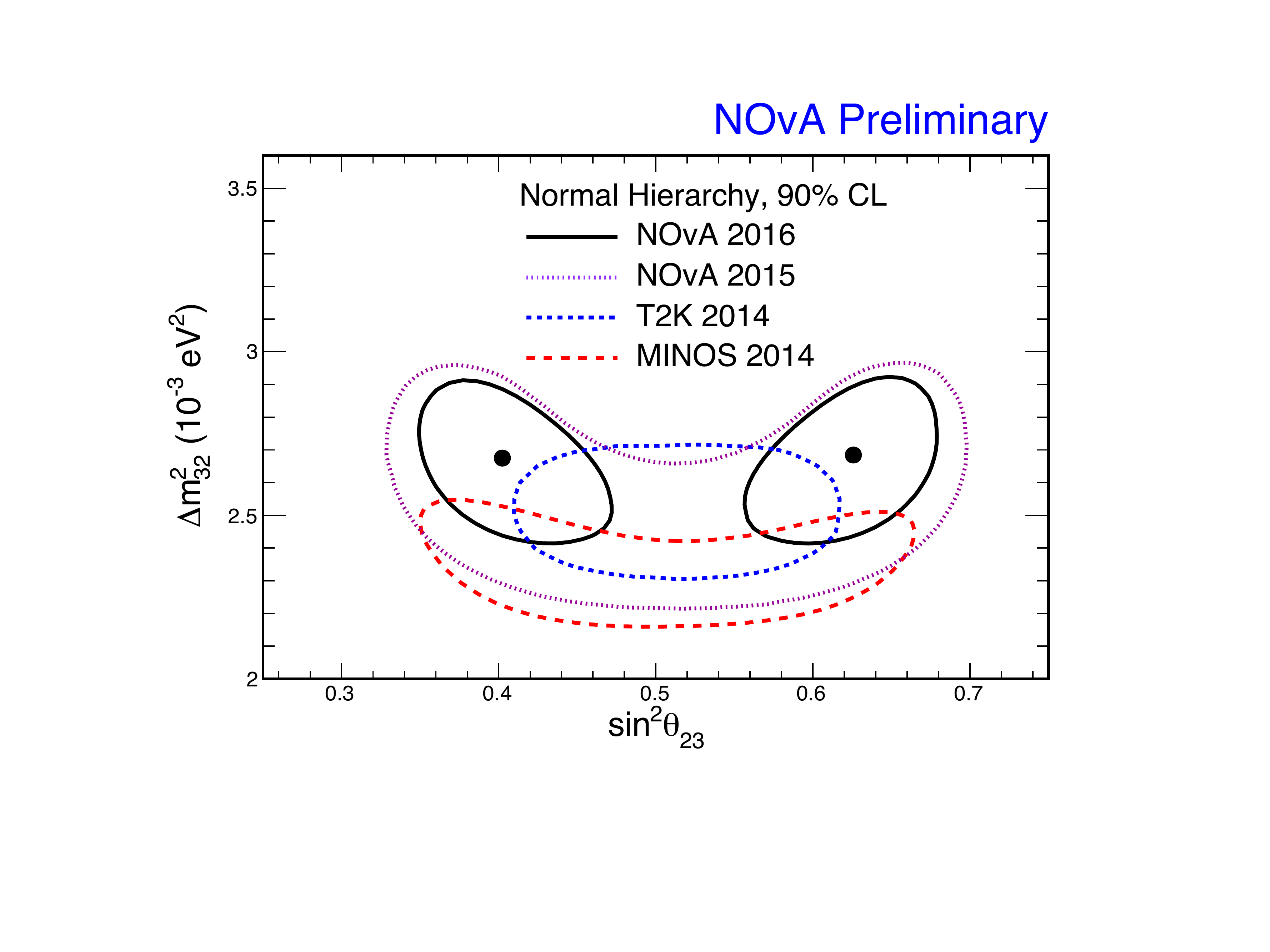}
  \end{center}
\caption{{\small The left plot compares the $\nu_{\mu}$-disappearance data to the best fit prediction and also show the unoscillated prediction one would have extected in
                 the absence of neutrino oscillations.
                 The middle plot shows the ratio of background subtracted oscillation to the unoscillated prediction.
                 The right plot shows the allowed 90\% C.L. regions in $\Delta m^{2}_{32}$ and $\sin^{2}\theta_{23}$ region.
                 The NOvA 2015 and 2016 results are compared with MINOS and T2K}}
\label{fig:numu_results}
\end{figure}

The $\nu_{\mu}$ disappearance data is fit for the $\Delta m^{2}_{32}$ and $\sin^{2}\theta_{23}$ parameters \cite{nova_numu_2016_results}.
Based on the predictions from the near detector in the absence of oscillation 473 $\nu_{\mu}$ events were expected in the FD
but we observe 78 events which is clear evidence of neutrino oscillations as shown in the left and middle plots of Fig. \ref{fig:numu_results}.
The best fit to the data gives $\Delta m^{2}_{32}$ = $(2.67 \pm 0.11) \times 10^{3} \text{eV}^{2}$ and the $\sin^{2}\theta_{23}$ at two statistically degenerate
values $0.404^{+0.030}_{-0.022}$ and $0.624^{+0.022}_{-0.030}$ both at the 68\% C.L. in the normal hierarchy (NH).
Maximal mixing, where $\sin^{2}\theta_{23}$ = 0.5, is disfavored by the data at 2.6 $\sigma$.
The right plot in Fig. \ref{fig:numu_results} shows the allowed 90\% C.L. regions in $\Delta m^{2}_{32}$ and $\sin^{2}\theta_{23}$.
The NOvA 2015 and 2016 results are compared with MINOS and T2K.
\subsection{$\nu_{e}$ FIT TO THE APPEARANCE DATA}
\begin{figure}[H]
\begin{center}
  \includegraphics[width=0.4\textwidth,height=0.20\textheight]{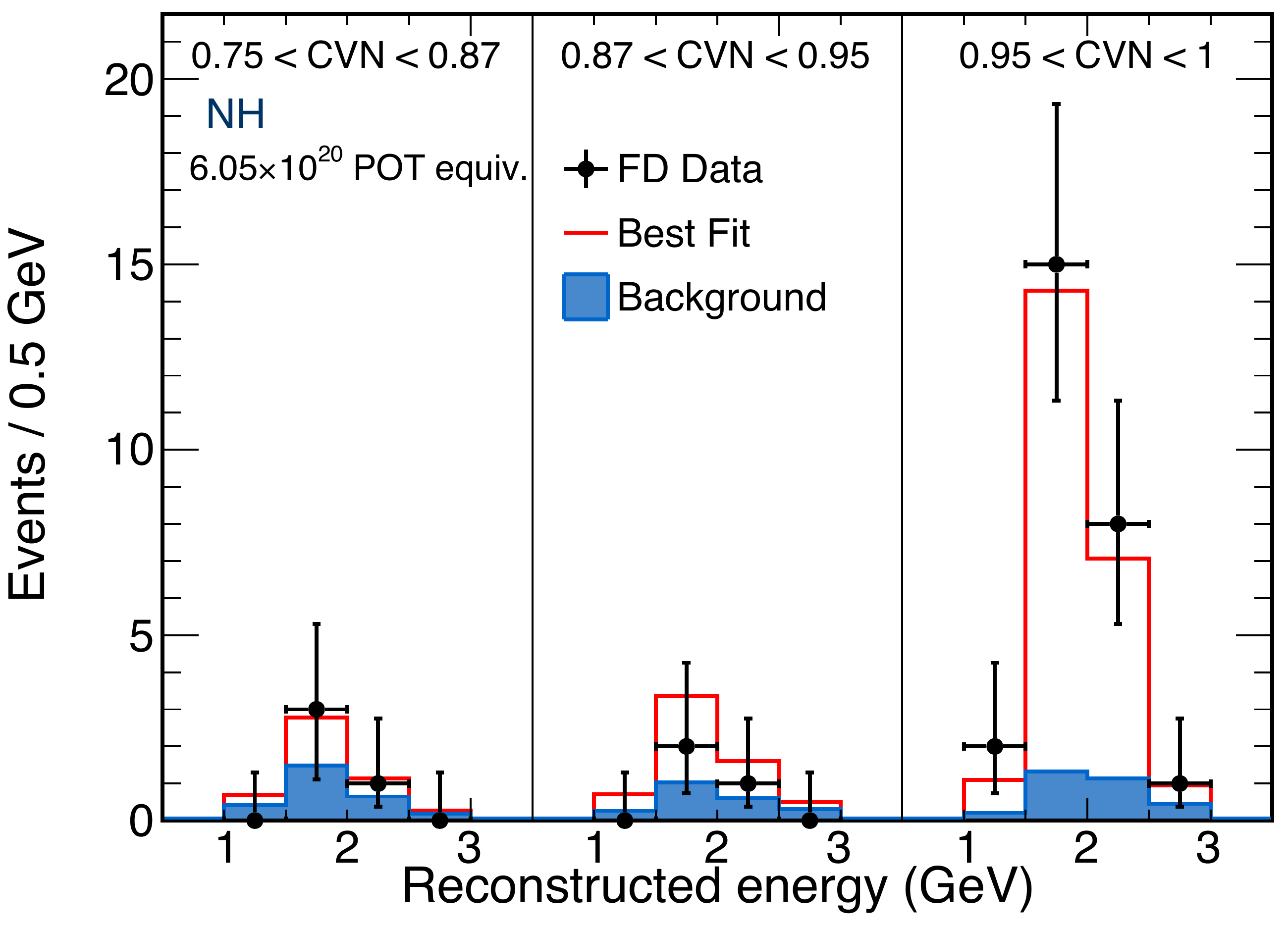}
  \includegraphics[width=0.4\textwidth,height=0.30\textheight]{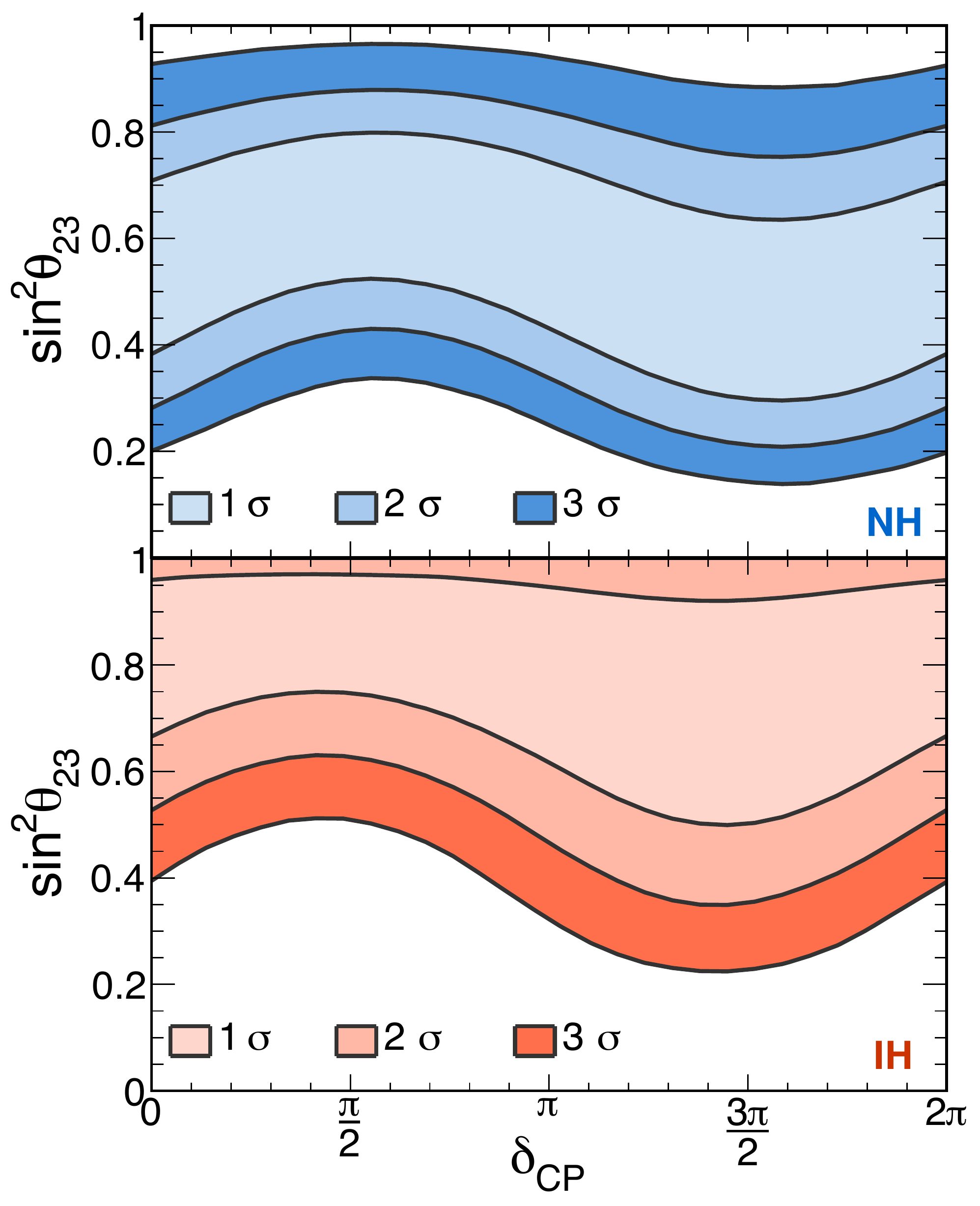}
  \end{center}
\caption{{\small The left plot shows a comparison of the event distribution with the expectations at the best-fit point as a function of the classifier variable and reconstructed neutrino energy.
                 The right plot shows the allowed regions of $\sin^{2}\theta_{23}$ and $\delta_{CP}$ parameters. The top panel is for normal hierarchy ($\Delta m^{2}_{32}$ $>$ 0)
                 and bottom panel is for inverted hierarchy ($\Delta m^{2}_{32}$ $<$ 0).}}
\label{fig:nue_results}
\end{figure}

The $\nu_{e}$ appearance data is fit for the $\sin^{2}\theta_{23}$ and $\delta_{CP}$ parameters.
The values $\sin^{2}2\theta_{13}$ = 0.085 $\pm$ 0.005 and
$\Delta m^{2}_{32}$ = +2.44 $\pm$ 0.06 (NH) are used as nuisance parameters in the $\nu_{e}$-appearance fit.
Thirty three, 33, electron neutrino candidates were observed with an expected background of 8.2$\pm$0.8 events; the significance of $\nu_{e}$ appearance is greater than 8 $\sigma$.
The left plot in Fig. \ref{fig:nue_results} shows a comparison of the event distribution with the expectations at the best-fit point as a function of the classifier variable and reconstructed neutrino energy.
The right plot in Fig. \ref{fig:nue_results} shows the allowed regions of $\sin^{2}\theta_{23}$ and $\delta_{CP}$ parameters for both normal and inverted hierarchies.
\subsection{SIMULTANEOUS FIT OF THE DISAPPEARANCE AND APPEARANCE DATA}
\begin{figure}[H]
\begin{center}
  \includegraphics[width=0.4\textwidth,height=0.30\textheight]{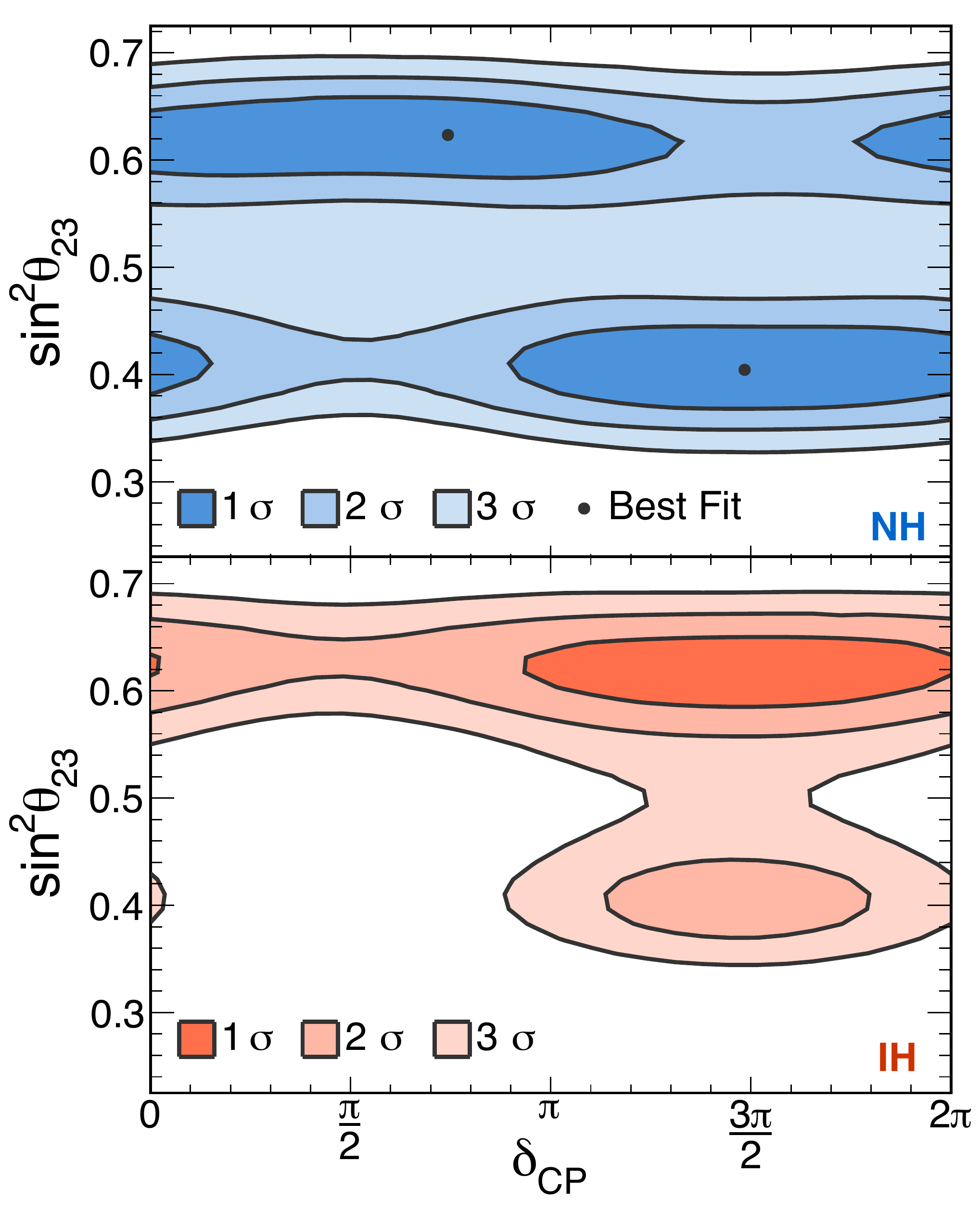}
  \end{center}
\caption{{\small The allowed regions of $\sin^{2}\theta_{23}$ and $\delta_{CP}$ parameters for the simultaneous fit of the NOvA's $\nu_{e}$ appearance and
         $\nu_{\mu}$ disappearance data.
         The top panel is for normal hierarchy ($\Delta m^{2}_{32}$ $>$ 0)
                 and bottom panel is for inverted hierarchy ($\Delta m^{2}_{32}$ $<$ 0). }}
\label{fig:numu_nue_contours}
\end{figure}

The constraints on the oscillation parameters can be improved by combining NOvA's $\nu_{e}$ appearance data with its $\nu_{\mu}$ disappearance data \cite{nova_numu_nue_2016_results}.
Appearance and disappearance data are simultaneous fit for the $\sin^{2}\theta_{23}$, $\delta_{CP}$ and $\Delta m^{2}_{32}$ parameters.
The parameter $\sin^{2}2\theta_{13}$ is 0.085 $\pm$ 0.005 a constraint from the reactor experiments \cite{reactor_dayabay, reactor_reno, reactor_pdg}.
Fig. \ref{fig:numu_nue_contours} shows the allowed regions of $\sin^{2}\theta_{23}$ and $\delta_{CP}$ parameters.
There are two degenerate best-fit points, both in the normal hierarchy $\sin^{2}\theta_{23}$ = $0.404$, $\delta_{CP} = 1.48\pi$
and $\sin^{2}\theta_{23}$ = $0.623$, $\delta_{CP} = 0.74\pi$.
The inverted mass hierarchy in the lower octant is disfavored at greater than 93\% C.L. for all values of $\delta_{CP}$ and excluded at greater than
3 $\sigma$ significance outside the range 0.97$\pi < \delta_{CP} < 1.94\pi$.
\section{ACKNOWLEDGMENTS}
This work was supported by the U.S. Department of Energy; the U.S. National Science Foundation; the Department of Science and Technology, India;
the European Research Council; the MSMT CR, GA UK, Czech Republic; the RAS, RMES, and RFBR, Russia; CNPq and FAPEG, Brazil; and the State and University of Minnesota. We are grateful for the contributions of the staffs at the University of Minnesota module assembly facility and Ash River Laboratory, Argonne National Laboratory, and Fermilab. Fermilab is operated by Fermi Research Alliance, LLC under Contract No. De-AC02- 07CH11359 with the U.S. DOE.



\begin{thebibliography}{}

\bibitem{nova_technical_report}NOvA Technical Design Report No. FERMILAB-DESIGN-2007-01.
\bibitem{nova_numi} P. Adamson et al., Nucl. Instrum. Methods Phys. Res., Sect. A 806, 279 (2016); NuMI Technical Design Handbook Report No. FERMILAB-DESIGN-1998-01.
\bibitem{nova_beam} Adamson P et al. 2016 Nucl. Instrum. Meth. A806 279
\bibitem{nova_apd}http://www.hamamatsu.com/us/en/product/alpha/S/4112/S8550‐02/index.html.
\bibitem{cvn_nue} C. Szegedy et al., arXiv:1409.4842.
\bibitem{nova_cvn} A. Aurisano, A. Radovic, D. Rocco, A. Himmel, M. D. Messier, E. Niner, G. Pawloski, F. Psihas, A. Sousa, and P. Vahle, J. Instrum. 11, P09001 (2016).
\bibitem{reactor_dayabay} F. P. An et al., Phys. Rev. Lett. 115, 111802 (2015).
\bibitem{reactor_reno} S. B. Kim, Nucl. Phys. B908, 94 (2016).
\bibitem{reactor_pdg} K. A. Olive et al. (Particle Data Group), Chin. Phys. C 38, 090001 (2014), and 2015 update.
\bibitem{nova_numu_2016_results} Measurement of the neutrino mixing angle $\theta_{23}$ in NOvA. NOvA Collaboration (P. Adamson (Fermilab) et al.).  Phys.Rev.Lett. 118 (2017) no.15, 151802. FERMILAB-PUB-17-019-ND
\bibitem{nova_numu_nue_2016_results} Constraints on Oscillation Parameters from $\nu_{e}$ Appearance and $\nu_{\mu}$ Disappearance in NOvA. NOvA Collaboration (P. Adamson (Fermilab) et al.). Phys.Rev.Lett. 118 (2017) no.15, 151802. FERMILAB-PUB-17-019-ND

\end{thebibliography}
\end{document}